# Use of Computer Applications for Determining the Best Possible Runway Orientation using Wind Rose Diagrams

Manmeet Singh[1] and Tanuj Chopra[2]

[1]Undergraduate Student, Department of Civil Engineering, Thapar University, Punjab, India
Email: manmeet20singh11@gmail.com

[2]Assistant Professor, Department of Civil Engineering, Thapar University, Punjab, India
Email: tanuj33@gmail.com

*Abstract*-**As technology advances, there are more uses of computer applications in a *Civil Engineering Projects*. To simplify the work, minimize errors, and obviate computational time required in designing and other structural analysis work, software which give accurate results to the given inputs are the need of the hour and this paper highlights the simulation of a manual procedure to find out the orientation of a Runway from the Wind Data available. A Computer application in *Java on Net beans platform* has been used to develop this Wind Rose Software. This software has been developed using the concepts of *Wind Rose Diagram Type-II* which will calculate the best possible orientation of the Runway along with the Coverage.**

*Index Terms*- **Orientation, Wind Rose, Netbeans IDE, Coverage, Cross Wind Component**

## I. INTRODUCTION

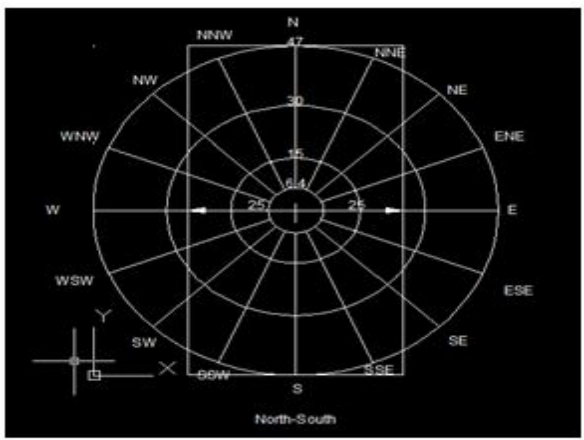

Figure 1. Wind Rose Diagram modelled in AutoCAD with a strip of 50kmph kept on it to calculate coverage

The orientation of a Runway is a very important parameter before the design and construction of an Airport. The concept behind deciding the direction is lucid one. The aircraft needs an uplifting force taking off and resisting force while landing to optimize and minimize the length of runway required and the energy consumption. Also the load on the engine is an important factor. So such an orientation needs to be found out and Type I and II Wind Rose Diagrams are used for the same. Type I is a simple one, but Type II is a rigorous one taking into consideration the intensities of Wind in addition to direction and the percentage of time, wherein wind blows in that particular direction, which are used in Type I. The Compass Rose is an old design element found on compasses, maps and even monuments to show cardinal directions (North, South, East and West) and even intermediate directions. The 'rose' term arises from fairly aesthetic figures used with early compasses. Today a Wind Rose is a graphic tool used by meteorologists to give a view of how wind speed and direction is distributed in a particular direction.

The orientation of a runway is found out by wind rose diagram type II .In this method three circles of intensities 6.4, 15, 30, and 47 kmph are drawn. As shown in the above diagram, the general method to calculate the Coverage corresponding to each direction by keeping a tracing paper of 50 kmph width. It is done by adding the percentages for each direction. The area for which the tracing paper cuts, are taken proportionately. The direction which has the maximum coverage is chosen as the runway direction. As per the specifications of FAA 95% coverage is required. So, if neither of the directions gives 95% coverage, the direction with maximum coverage is chosen and another runway perpendicular to the above is selected as the second option to compensate for the Coverage required.

## II. DEVELOPMENT OF ALGORITHM TO MAKE COMPUTER APPLICATION

As has been explained in the upper text, tracing paper with 50 kmph width is kept in all directions and coverage calculated by summing up the corresponding contribution by each area which the tracing paper overlaps. Three arrays for the coefficients to be multiplied were declared and it was seen that, in each iteration, multiplying the area by the coefficient we get the tantamount effect as was done manually on a graph sheet or approximately. For calculating the coefficients AutoCAD drawings of each iteration was made and the coefficients calculated by dividing the overlapping area with the contributing area. It was seen, for example, that the coefficients for north south direction when the tracing paper was kept in the north south were identical to all coefficients of any other direction, but now the coefficients changed their loyalties .Three arrays.

## III. Coding The Method Developed

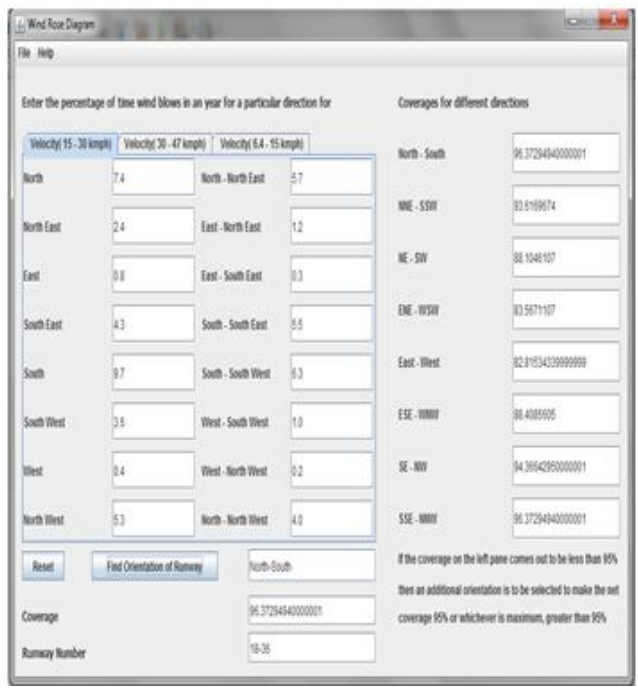

Figure 2. Application after running with a sample data

Netbeans IDE 7.1.1 was used for coding the above developed algorithm into a fully functional Graphical User Interface Application. Basic Java Programming was used to code the processThe window below shows the interface developed. It was generated by declaring a number of text fields, labels, a frame, two menu bars with menu items, a Tabbed pane, and lastly buttons. Figure 2. [1]GUI for simulating the Wind Rose Diagram Method

```
public WindRoseDiagram() {
    initComponents();
    textfields = new JTextField[][]{
        {EFf415, NNEFf415, NEFf415, ENEFf415, EFf415, ESEFf415, SEFf415, SSEFf415, SFf415, SSWFf415, SWFf415, WSW
        ...
    };
```

Figure 3. Two Dimensional array made out of entered values in textfields

coeff1[8], coeff2[8], and coeff3[8] were declared for velocities 6.4-15, 15-30 and 30-47 kmph. And each element of the array represented each direction. So now if calculate coefficients for a particular direction say North South and calculate its coefficients and add the percentages multiplied by coefficients, we'll get the coverage for North South. As per the observation discussed above, the coefficient array was shifted one position to the right accounting for the change in direction of the tracing paper and again coverage calculated. An important observation here is to note down the naming style of the textfields. The alphabets before F denote the direction, after F the particular velocity and F denoting the text field. Similarly the labels were named but without the use of the letter F. Now the last task in coding was to assign events related to buttons. For assigning the events related to buttons, in the Design View the respective button was right clicked with the help of mouse and pointed to Action then actionPerformed[variable name of button] was clicked. This assigns an event to a button and the event can be written in the respective function.

## IV. Assigning Events

### A. Reset Button

```
809    private void resetButtonActionPerformed(java.awt.event.ActionEvent evt)
810        for (JTextField rows[] : textFields) {
811            for (JTextField tf : rows) {
812                tf.setText("0");
813            }
814        }
815    }
```

The above code conceives a loop which goes on to assign each element of the 2D array textFields as "0".

### B. Finding Orientation Of Runway

This was the main action performed in the application and all the data input by the user in the textfields was used here. As can be seen on the image on the next page first of all, all the textfields data was transferred to three 1D array with the help of the functions parseDouble and getText. For this three 1D arrays vel1[], vel2[] and vel3[] were declared of the size equal to the number of directions i.e. 8. A number of integer variables i, j, sum, arraySize, calmPeriod, temp1 and temp2 were declared initially, which were just used to perform various computations. I and j were used to run various single and nested loops. Sum and calmPeriod were used to calculate the calm period (velocity less than 6.4kmph), which is added in the end to all the coverages as it contributes to all directions. arraySize is utilized to decide the number of elements in the vel[] arrays and is just used to declare in an aesthetic way rather than declaring otherwise.temp1 and temp2 are used to shift the coefficient matrices one position forward while calculating coverage for each iteration. Its concept lies in the fact as was explained in the initial stages, that the coefficients to be multiplied to a particular percentage, to account for the partially overlapping area of the Wind Rose Diagram, are the same if they are shifted to the right by one position to calculate the coverage of a particular direction. add[] array was declared and used to calculate the coverage excluding the Calm Period. With the help of a nested loop of I and j the elements of the array add were calculated. Initial value of the array was set as 0 and in an internal j loop the add elements calculated by adding the initial value of add and summation of multiplied velocities with their respective coefficients. After moving out of the j loop once and calculating a single add element, the coefficients are shifted by one position by an algorithm as shown above and the next iteration of I loop calculates next add element. This goes on upto eight iterations and all add elements are calculated. Next by running a for loop all the add elements are increased by an amount equal to Calm Period.Using setText(String.valueOf(Variable whose value is to be set in textfield)) function all the text fields corresponding to the different coverages are assigned the values of their respective coverages.

```
private void findOrientationButtonActionPerformed(java.awt.event.ActionEvent evt) {
    int i, n, arraySize = 8;
    double sum = 0;
    double temp1, temp2;
    double CalmPeriod;

    double add[] = new double[arraySize];
    double coff1[] = {1, 1, 1, 0.831353, 0.626081, 0.831353, 1, 1};
    double coff2[] = {1, 1, 0.358123, 0, 0, 0.358123, 1 };
    double vel1[] = new double[arraySize];
    double vel2[] = new double[arraySize];
    double vel3[] = new double[arraySize];
    double coverage[] =new double[8];
    for (int j = 0; j < vel1.length; j++) {
        vel1[j] = Double.parseDouble(textFields[0][j].getText());
    }
    for (int j = 0; j < vel2.length; j++) {
        vel2[j] = Double.parseDouble(textFields[1][j].getText());
    }
    for (int j = 0; j < vel3.length; j++) {
        vel3[j] = Double.parseDouble(textFields[2][j].getText());
    }

    for (i = 0; i < 8; i++) {
        sum = sum + vel1[i] + vel2[i] + vel3[i];
    }
    CalmPeriod = 100 - sum;
```

```
    for (i = 0; i < 8; i++) {
        for (int j = 0; j < 8; j++) {
            add[i] = add[i] + coff1[j] * vel2[j] + coff2[j] * vel3[j];
        }
        temp1 = coff1[7];
        temp2 = coff2[7];
        for (int k = 7; k >= 1; k--) {
            coff1[k] = coff1[k - 1];
            coff2[k] = coff2[k - 1];
            coff1[0] = temp1;
            coff2[0] = temp2;
        }
    }
    for(i=0;i<8;i++)
    {
        coverage[i] =(CalmPeriod + add[i]);
    }
    double max = coverage[0];
    n = 0;
    for (i = 0; i < 8; i++) {
        if (coverage[i] > max) {
            max = coverage[i];
            n = i;
        }
    }
    NorthSouthF.setText(String.valueOf(coverage[0]));
    NNESSWF.setText(String.valueOf(coverage[1]));
    NESWF.setText(String.valueOf(coverage[2]));
    SNWNSEF.setText(String.valueOf(coverage[3]));
    EastWestF.setText(String.valueOf(coverage[4]));
    NSESNWF.setText(String.valueOf(coverage[5]));
```

```
    NSESNWF.setText(String.valueOf(coverage[5]));
    SENWF.setText(String.valueOf(coverage[6]));
    SSENNWF.setText(String.valueOf(coverage[7]));
    if (max<=100)
    {
        switch (n) {
            case 0:
                ResultF.setText("North-South");
                RunwayNumberF.setText("18-36");
                CoverageF.setText(String.valueOf(coverage[0]));
                break;
            case 1:
                ResultF.setText("NNE-SSW");
                RunwayNumberF.setText("01-19");
                CoverageF.setText(String.valueOf(coverage[1]));
                break;
            case 2:
                ResultF.setText("NW-SE");
                RunwayNumberF.setText("02-20");
                CoverageF.setText(String.valueOf(coverage[2]));
                break;
            case 3:
                ResultF.setText("SNW-NSE");
                RunwayNumberF.setText("03-21");
                CoverageF.setText(String.valueOf(coverage[3]));
                break;
            case 4:
                ResultF.setText("East-West");
                RunwayNumberF.setText("04-22");
                CoverageF.setText(String.valueOf(coverage[4]));
                break;
```

As shown in the above image, as our concept says that the direction having maximum coverage will be chosen as the runway direction, using the switch case command the case is found out which has maximum coverage. For this while calculating the maximum coverage value a variable n is assigned the value of the position of maximum coverage in the 1D array coverage[]. In the switch case the text field corresponding to the final orientation of runway, runway number and its coverage fields are assigned. The whole switch case is enclosed in an "if else" command which allows the control to enter switch only if the summation of the values input by the user are less than 100. If so then the required result is obtained, else the field corresponding to the final orientation of runway shows a message, "Coverage can't be greater than 100" giving the user an idea that erroneous values have been entered or recorded.

```
                break;
            case 5:
                ResultF.setText("NSE-SNW");
                RunwayNumberF.setText("05-23");
                CoverageF.setText(String.valueOf(coverage[5]));
                break;
            case 6:
                ResultF.setText("SE-NW");
                RunwayNumberF.setText("06-24");
                CoverageF.setText(String.valueOf(coverage[6]));
                break;
            case 7:
                ResultF.setText("SSE-NNW");
                RunwayNumberF.setText("07-25");
                CoverageF.setText(String.valueOf(coverage[7]));
                break;

        }
    }
    else
    {
        CoverageF.setText("Coverage can't be greater than 100");
    }
}
```

## C. Exit Button

```
797    private void exitMenuItemActionPerformed(java.awt.event.ActionEvent evt)
798        System.exit(0);
799    }
```

The event is associated with the exit button in an analogous fashion as was with the other buttons explained above. System.exit(0) function is used to close the application if the user clicks on the exit button at the bottom or the one in the file menu.

## D. Menu Items

File Menu has only the exit menu item and its function is the same as was for Exit button.Help Menu has two Menu items.

## E. About Menu Item

This Menu Item was designed as a dialog box which becomes active when the About menu item is put into action. The dialog box was created as a separate class and then an object of that class declared in the main class and used appropriately.

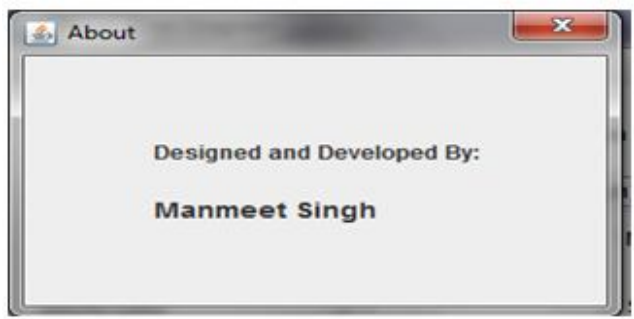

As can be seen from the snap below an object dlg of AboutDialog class is declared and its visibility is set to true, which means that while it would be active the back window of main program remains in dormant state. As the Grid Layout was used previously here Border layout is used to make the border of the dialog box visible. And then a label added to the Dialog Box to show the message.

```
809    private void aboutMenuItemActionPerformed(java.awt.event.ActionEvent evt) {
810        AboutDialog dlg = new AboutDialog(this, true);
811        dlg.setVisible(true);
812    }
```

## V. HELP MENU ITEM

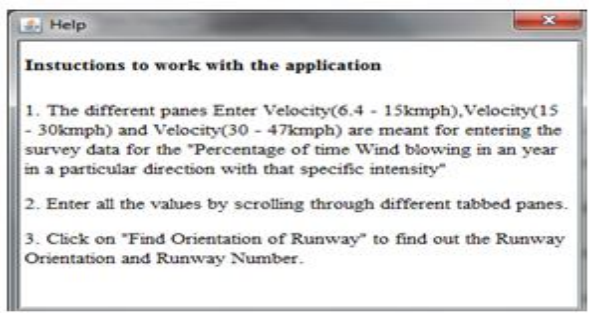

This menu item is made similar to the about menu item but with slight changes since the message is longer.

```
6    <html>
7      <head>
8      </head>
9      <body>
10       <h3>Instructions to work with the application</h3>
11       <p>1. The different panes Enter Velocity(6.4 - 15kmph),Velocity(15 - 30kmph)
12       and Velocity(30 - 47kmph) are meant for entering the survey data for the
13       "Percentage of time Wind blowing in an year in a particular direction into
14       that specific intensity"</p>
15       <p>2. Enter all the values by scrolling through different tabbed panes.</p>
16       <p>3. Click on "Find Orientation of Runway" to find out the Runway Orientation and
17       Runway Number.</p>
18      </body>
19    </html>
```

An HTML file htmlhelp.html is made and the following code above is written into it. Now this HTML file would be read by an Editor pane in the Dialog Box of help (Dialog box created as was in About), and then an object of Help class made in the main class. Whenever a call is made to the Help Menu item a dialog box with the message written with the help of HTML is displayed.

```
814    private void helpMenuItemActionPerformed(java.awt.event.ActionEvent evt)
815        HelpDialog dlg = new HelpDialog(this, false);
816        dlg.setVisible(true);
817    }
```

Above code is similar to About Menu item code except that the visibility is set to false. So the main window screen would also remain active when the help is active. Shown below the code written to read the message by HelpDialog class from the HTML file.First of all the libraries FileReader, InputStream and HTMLEditorKit are imported to the class to use their functions in the class and the file htmlhelp.html read using the function get ResourceAsStream.

```
4    */
5
6    package windrosediagram;
7    import java.io.FileReader;
8    import java.io.InputStream;
9    import javax.swing.text.html.HTMLEditorKit;
10
11   /**
12    *
13    * @author paramjeet singh
14    */
15   public class HelpDialog extends javax.swing.JDialog {
16
17       /**
18        * Creates new form HelpDialog
19        */
20       public HelpDialog(java.awt.Frame parent, boolean modal) {
21           super(parent, modal);
22           initComponents();
23           try {
24               jEditorPane1.setEditorKit(new HTMLEditorKit());
25               InputStream input = this.getClass().getResourceAsStream("htmlhelp.html");
26               jEditorPane1.read(input, this);
27           } catch (Exception e) {
28               e.printStackTrace();
29           }
30       }
31   }
```

Finally the message which guides the user to chose a combination of Orientations if all coverages are less than 95%, was written by inserting three labels and then editing their texts.

If the coverage on the left pane comes out to be less than 95%

then an additional orientation is to be selected to make the net

coverage 95% or whichever is maximum, greater than 95%

## VI. Conclusions

The Wind Rose method can be accurately simulated by using programming in an effective way. The application was checked with the example 6.1 on page 70 of reference [2] and found to work with good level of precision, with the results matching upto fourth place of decimal. Thus the application can be satisfactorily used for the finding the orientation of a Runway.